\documentstyle[aps,preprint]{revtex}

\begin{document}


\tightenlines


\title{Dislocation scattering in a two dimensional electron gas}

\author{Debdeep Jena \thanks{djena@indy.ece.ucsb.edu},Arthur. C. 
Gossard, Umesh. K. Mishra}
\address{Department of Electrical and Computer Engineering, University
of California, Santa Barbara, CA 93106 }

\maketitle


\begin{abstract}

A theory of scattering by charged dislocation lines in a two-dimensional electron gas (2DEG)
is developed.  The theory is directed towards understanding transport in AlGaN/GaN 
high-electron-mobility transistors (HEMT), which have a large number of line dislocations 
piercing through the 2DEG.  The scattering time due to dislocations is derived for a 2DEG 
in closed form.  This work identifies dislocation scattering as a  mobility-limiting 
scattering mechanism in 2DEGs with high dislocation densities.  The insensitivity of the 
2DEG (as compared to bulk) to dislocation scattering is explained by the theory.

\end{abstract}

\pagebreak


The last decade has witnessed a rapid development of III-V materials for a variety of 
optoelectronic and high power high-speed electronic devices [Ref 1-3].  The wide range 
of bandgaps available by alloying GaN with Al and In have made it suitable for 
high-power, high-voltage electronic applications.  AlGaN/GaN heterostructures have 
been an area of active research, owing to the demonstration of high power microwave 
high electron mobility transistors (HEMTs) [Ref 3].  A good substrate, however, still 
remains elusive.  Due to the large lattice mismatch with the present substrate of choice, 
sapphire, state of the art AlGaN/GaN HEMTs have [Ref 4-5] a two-dimensional-electron-gas 
(2DEG) which has $1-100X10^{8}/cm^2$ line dislocations passing through it.  Efforts to improve mobilities 
by reducing the number of dislocations have resulted in the novel growth technique of 
lateral epitaxial overgrowth [LEO] [Ref 6-7].  Recently, Look and Sizelove [Ref 8] 
analyzed the effect of dislocation scattering on the mobility of 3D bulk GaN.  However, 
efforts[Ref 9-11] at explaining the observed mobilities in AlGaN/GaN 2DEGs have not 
considered scattering by dislocations.  Surprisingly, there exists no adequate theory 
for dislocation scattering in a 2DEG.  An effort was made to treat dislocation scattering 
in a AlGaAs/InGaAs/AlGaAs quantum well (QW) before the advent of AlGaN/GaN heterostructures 
[Ref 12].  We point out the limitations of the treatment.  This paper develops the theory 
for the effect of dislocation scattering on transport in 2DEGs.  The theory developed shows 
that 2DEG mobility is affected strongly by a high density of dislocations.  The effect is 
weaker, however, than that in 3D bulk; the reasons for this are pointed out.

The system we consider is a perfect 2DEG, i.e., there is no spatial extent of carriers in the 
growth direction.  The electrostatic potential $\phi$ satisfies Poisson's equation, which for the 
2DEG under consideration with the Thomas Fermi approximation is [Ref 13] 

\begin{equation} 
\nabla^{2} \phi - 2q_{TF}\phi ({\bf r})\delta (z) = 
			-\frac{4\pi \rho_{ext}}{\epsilon_{0} \epsilon_{b}}
\end{equation}

Here $\rho_{ext}$ is the external charge, $\epsilon_{0} \epsilon_{b}$ is the dielectric 
constant in the material (assumed to be 
same in the 2DEG and outside it), ${\bf r}=(x,y)$ is the in-plane vector, 
and $q_{TF}=\frac{2}{a_{B}^{*}}$ is the 2D Thomas Fermi 
wavevector, $a_{B}^{*}$ being the effective Bohr radius in the material.  
A valley degeneracy of one and spin degeneracy of two is implied.  The conventional 
technique to arrive at the 
screened potential is to use the Fourier-Bessel expansion of the potential
$\phi({\bf r},z)=\int_{0}^{\infty} q A_{q}(z) J_{0}(qr) dq$
, where $J_{0}$ is 
the Bessel function of order zero, and $A_{q} (z)$ is the Fourier-Bessel coefficient,
$q$ being the 
in-plane wavevector.  The value of the coefficient averaged over zero spatial extent 
along the $z$ axis is $A(q)=A_{q}(0)$.  The Fourier Transform of the screened 2D 
potential due to a point charge at a distance $z$ from the plane of the perfect 
2DEG is given by [Ref 13-14]

\begin{equation}
A(q)= \frac{e^2}{2 \epsilon_{0} \epsilon_{b}} \cdot 
        \frac{e^{-q|z|}}{q+q_{TF}}
\end{equation}

where $e$ is the electronic charge.  The previous effort mentioned for quantum well 
dislocation scattering [Ref 12]  a) arrives at a scattering potential which models an 
in-plane charged impurity rather than the spatially extending dislocation line, and  
b) does not consider the strong screening contribution in the highly degenerate 2DEG.  
We take these important factors into consideration.

A line dislocation has been shown [Ref 8,15,16] to be accurately modeled by a line of charge.  
Treating the line dislocation as a line of charge with charge density $\rho_{L}$ [Fig 1], 
we write 
the scattering contribution by a elementary length $dz$ of the line charge to the screening 
potential as

\begin{equation}
dA(q)=\frac{e}{2 \epsilon_{0} \epsilon_{b}} \cdot 
        \frac{dz \rho_{L} e^{-q|z|}}{q+q_{TF}}
\end{equation}

Integrating over an infinite length of the line charge, we get the new screened potential as

\begin{equation}
A(q)=\frac{e}{2 \epsilon_{0} \epsilon_{b}} \cdot 
        \frac{2 \rho_{L}}{q(q+q_{TF})}
\end{equation}

There are $N_{dis}$ line dislocations piercing the 2DEG per unit area.  The scattering 
rate for a degenerate 2DEG is given by [Ref 14]

\begin{equation}
\frac{1}{\tau_{dis}^{2D}} =
	N_{dis} \cdot (\frac{m^{*}}{2 \pi \hbar^3 k_{F}^{3}}) \cdot
	   \int_{0}^{2k_{F}} |A(q)|^{2} \frac{q^2 dq}{\sqrt{1-(\frac{q}{2k_{F}})^2}}
\end{equation}

where $k_{F}=\sqrt{2 \pi n_{s}}$ is the Fermi wavevector [Ref 13], which depends on the 
2DEG carrier concentration $n_{s}$.  
Using the screened potential [Eq 4] with the substitution $u=\frac{q}{2k_{F}}$ we get 
the scattering rate as 

\begin{equation}
\frac{1}{\tau_{dis}^{2D}} =
	\frac{N_{dis} m^* e^{2} \rho_{L} }{\hbar^{3} \epsilon_{0}^{2} \epsilon_{b}^{2}} 
\cdot (\frac{1}{16 \pi k_{F}^{4}}) 
\cdot \int_{0}^{1} \frac{du}{(u+\frac{q_{TF}}{2k_{F}})^{2} \sqrt{1-u^2}}
\end{equation}

The dimensionless integral $I(\frac{q_{TF}}{2k_{F}})$ can be evaluated exactly.  It 
depends only on the 2DEG carrier 
concentration through the Fermi wavevector, and has an approximate $\sqrt{n_{s}}$ 
dependence.  [Fig 2] 
shows it's exact dependence on the 2DEG carrier concentration for $\epsilon_{b}=10$ and 
$m^{*}=0.228m_{0}$ (for GaN).  The 
integral is closely approximated by the relation $A \cdot (\frac{2k_{F}}{q_{TF}}) + B$, 
where $A$ and $B$ were determined to be 
$0.92$ and $-0.25$ respectively.  $\rho_{L}$, the line charge density is given to a very good 
approximation [Ref 8,15] by $\frac{e \cdot f}{c_{0}}$, where $c_{0}$ is the lattice spacing in 
the $(0001)$ direction of 
wurtzite GaN, and $f$ is the fraction of filled states, calculated by Weimann et. al.  Using 
this, the 2D dislocation scattering time is

\begin{equation}
\tau_{dis}^{2D} =
	\frac{\hbar^{3} \epsilon_{0}^{2} \epsilon_{b}^{2} c_{0}^{2} }{N_{dis} m^* e^{4} f^{2} } 
\cdot \frac{16 \pi k_{F}^{4}}{I(\frac{q_{TF}}{2k_{F}})}  \approx
	\frac{\hbar^{3} \epsilon_{0}^{2} \epsilon_{b}^{2} c_{0}^{2} }{N_{dis} m^* e^{4} f^{2} } 
\cdot \frac{16 \pi k_{F}^{4}}{(\frac{1.84 k_{F}}{q_{TF}}-0.25)}
\end{equation} 

The scattering time has an approximate $n_{s}^{\frac{3}{2}}$ dependence.

The scattering time arrived at highlights the metallic nature of the 2DEG electrons.  The 
screening length for a 2DEG depends on $q_{TF}$ and $k_{F}$.  The Thomas Fermi wavevector 
$q_{TF}$ is constant.  
As the free carrier density is increased, $k_{F}$ increases, and $\lambda_{F}$, the Fermi 
wavelength gets 
shorter, leading to better screening.  The 2DEG carrier density does not freeze out at low 
temperatures as in 3D.  These factors contribute to the observed high mobilities in a 2DEG.  
In contrast, 3D screening and scattering is controlled by the Debye screening factor [Ref 14]
,$q_{D}=\sqrt{\frac{e^{2} n^{'}}{\epsilon k_{B} T}}$, where $n^{'}$ is the effective screening 
concentration, involving both free and bound carriers.  
At low temperatures, free carriers freeze out exponentially in a semiconductor.  An elongation 
of the Debye screening length $\lambda_{D}=\frac{1}{q_{D}}$ leads to a weaker screening.  
In addition, the carriers are 
less energetic, leading to strong scattering, and hence to lower mobilities.  A comparison of 
the dislocation scattering time in 2D [Eq 7] with the 3D bulk scattering time calculated 
recently by Look and Sizelove [Ref 8] (and also by  Pödör [Ref 17]) bears out the discussion 
above.  The result for 3D scattering time for $f$ fractionally filled states is

\begin{equation}
\tau_{dis}^{3D}=\frac{\hbar^{3} \epsilon_{0}^{2} \epsilon_{b}^{2} c_{0}^{2}}
			{N_{dis} m^{*} e^{4} f^{2}} 
\cdot
		\frac{(1+4\lambda_{D}^{2} k_{\perp}^{2})^{\frac{3}{2}}}{\lambda_{D}^{4}} 
\propto
			\frac{(k_{B} T)^{\frac{3}{2}}}{\lambda_{D}}
\end{equation}

where $k_{\perp}$ is the wavevector for electron motion perpendicular to the dislocations,
$k_{B}$ is the 
Boltzmann constant and $T$ is the temperature.

2DEG mobility inhibited by dislocation scattering only , given by $\mu_{dis}^{2D}$ is plotted 
in [Fig 3].  
Here we have assumed $f=1$, i.e. that all the acceptor states in the dislocation are filled.  
Thus the mobility calculated is for the worst case.  The dependence of mobility on 2DEG 
sheet density and dislocation density is given to an extremely good approximation by 
$\mu_{dis}^{2D} \propto \frac{n_{s}^{\frac{3}{2}}}{N_{dis}}$.

Our theory shows that for dislocation density of $10^{10}/cm^{2}$ and carrier densities 
in the $10^{12}-10^{13}/cm^{2}$ range, 
maximum 2DEG mobilities will be in the $10^{3}-10^{4} \frac{cm^{2}}{V \cdot s}$ range.  
A reduction in the dislocation density 
to $\approx 10^{8}/cm^{2}$ in AlGaN/GaN HEMTs has resulted in record high mobilities at 
low temperatures [Ref 18,19].  
The record 2DEG mobility as of today stands at $51,700 \frac{cm^{2}}{V \cdot s}$, observed 
at $13K$ with a sheet carrier density 
of $2.23X10^{12}/cm^{2}$.  By comparison, experiment and theory for a 3DEG at $13K$ and 
carrier concentration $10^{18} /cm^{3}$   
gives $\mu \approx 100 \frac{cm^{2}}{V \cdot s}$.  The relative insensitivity of a 2DEG 
to dislocation scattering as compared to 3D 
bulk leads us to propose that modulation doped structures should be considered for replacing 
bulk GaN in structures infected with high dislocation densities.

In conclusion, we have developed a theory for dislocation scattering in semiconductor 
heterostructure 2DEGs.  The theory explains the observed low temperature mobility enhancement  
in AlGaN/GaN HEMTs upon reduction of dislocation density.  The theory finds a simple parallel 
to the 3D scattering theory for line dislocations, and highlights the strong screening effect 
by carriers in a 2DEG which helps it achieve much higher mobilities than 3D bulk values.


\section*{Acknowledgments}

The authors would like to acknowledge helpful discussions with 
B. Heying, C. Elsass, I. Smorchkova, P. Chavarkar , J. Singh, and 
J. Speck.



\begin{figure}
\caption{Line dislocation modeled as a line of charge.  The dislocation line has filled 
acceptor states along it.  Charges on the dislocation line act as an extended remote impurity.  
The remote ionized impurity matrix element is integrated for all the small elemental remote 
impurities to account for the effect of the whole dislocation.
}
\label{Figure 1}
\end{figure}

\begin{figure}
\caption{Integral factor vs 2DEG carrier density.  The dimensionless integral in Eq. 6 
is evaluated exactly.  The plot here shows it's dependence on the 2DEG carrier density for
 $\epsilon = 10$ and $m^{*}=0.228m_{0}$.  The approximate $\sqrt{n_{s}}$ dependence is clearly 
seen.}
\label{Figure 2}
\end{figure}

\begin{figure}
\caption{Dislocation scattering inhibited 2DEG mobility.  Mobilities for three different 
dislocation densities have been shown.  Our theory predicts an approximate 
$\frac{n_{s}^{\frac{3}{2}}}{N_{dis}}$ mobility dependence, which is seen in this plot.  
Note the strong mobility degradation at $10^{10}/cm^{2}$ dislocation density.
}
\label{Figure 3}
\end{figure}



\begin{thebibliography}{99}

\bibitem{[1]} S. Nakamura, M. Senoh, N. Isawa, S. I. Nagahama, T. Yamada, T. Matshushita, Y. Sigimto, and H. Kikoyu, Appl. Phys. Lett.  70, 1417 (1997)
\bibitem{[2]} H. Sakai, T. Takeuchi, S. Sota, M. Kasturagawa, M. Komori, H. Amano, and I. Akasaki, J. Cryst. Growth  189/190, 831 (1998)
\bibitem{[3]} Y. F. Wu, B. P. Keller, P. Fini, S. Keller, T. J. Jenkins, L. T. Kehias, S. P. DenBaars, and U. K. Mishra, IEEE Electron Device Lett.  19, 50 (1998)  
\bibitem{[4]} D. Kapolnek, X. H. Wu, B. Heying, S. Keller, B. P. Keller, U. K. Mishra, S. P. DenBaars, and J. S. Speck, Appl. Phys. Lett.  67, 1541 (1995)
\bibitem{[5]} S. D. Lester, F. A. Ponce, M. G. Craford, and D. A. Steigerwald, Appl. Phys. Lett.  66, 1249 (1995)
\bibitem{[6]} K.Kato, T. Kusunoki, C. Takenaka, T. Tanahashi, and K. Nakajima, J. Cryst. Growth  115, 174 (1991)
\bibitem{[7]} H. Marchand, X. H. Wu, J. P. Ibbetson, P. T. Fini, P. Kozodoy, S. Keller, J. S. Speck, S. P. DenBaars, and U. K. Mishra, Appl. Phys. Lett.  73, 747 (1998)
\bibitem{[8]} D. C. Look and J. R. Sizelove, Phys. Rev. Lett.  82, 1237  (1999) 
\bibitem{[9]} M. Shur, B. Gelmont, and M. A. Khan, J. Electronic Materials 25, 5, 777 (1996)
\bibitem{[10]} Y. Zhang and J. Singh, J. Appl. Phys.  85, 587  (1999)
\bibitem{[11]} L. Hsu, and W. Walukiewich, Phys. Rev. B  56, 1520  (1997)
\bibitem{[12]} D. Zhao, and K. J. Kuhn, IEEE Trans. Electronic Devices  38, 2582 (1991)
\bibitem{[13]} T. Ando, A. B. Fowler and F. Stern, Rev. Mod. Phys.  54, 437  (1982)
\bibitem{[14]} J. H. Davies, The Physics of Low Dimensional Semiconductors, Cambridge University Press (1998)
\bibitem{[15]} N. G. Weimann, L. F. Eastman, D. Doppalapudi, H. M. Ng and T. D. Moustakas, J. Appl. Phys.  83, 3656  (1998)
\bibitem{[16]} H. M. Ng, D. Doppalapudi, T. D. Moustakas, N. G. Weimann and L. F. Eastman, Appl. Phys. Lett.  73, 821 (1998)
\bibitem{[17]} B. Pödör, Phys. Status Solidi, 16, K167 (1966)
\bibitem{[18]} C. R. Elsass, I. P. Smorchkova, B. Heying, E. Haus, P. Fini, K. Maranowski, J. P. Ibbetson, S. Keller, P. M. Petroff, S. P. Denbaars, U. K. Mishra and  J. S. Speck, Appl. Phys. Lett. 74, 3528 (1999)
\bibitem{[19]} I. P. Smorchkova, C. R. Elsass, J. P. Ibbetson, R. Vetury, B. Heying, P. Fini, E. Haus, S. P. DenBaars, J. S. Speck, and U. K. Mishra , J. Appl. Phys.  86, 4520 (1999)

\end{thebibliography}
\end{document}